\documentclass[12pt]{article}
\usepackage{fullpage}
\usepackage[authoryear]{natbib}
\usepackage{setspace}

\usepackage[utf8]{inputenc}
\usepackage{amsmath}
\usepackage{amssymb}
\usepackage{amsfonts,amsthm}
\usepackage{bbm}
\usepackage{graphicx}
\usepackage{float}
\usepackage{caption}
\usepackage{subcaption,mathrsfs,multirow}
\usepackage[ruled]{algorithm2e}
\usepackage{array}
\usepackage{booktabs}
\usepackage{authblk}
\usepackage{xr}
\usepackage{threeparttable}

\usepackage{verbatim}
\usepackage{listings}
\usepackage{color}

\usepackage{epsfig}
\usepackage{wasysym}

\newtheorem{assumption}{Assumption}
\newtheorem{definition}{Definition}
\newtheorem{theorem}{Theorem}

\newtheorem{proposition}{Proposition}

\usepackage{color}

\newcommand{\comout}[1]{}

\def\lpbox#1{\vskip1mm \begin{center}
        \hspace{.0\textwidth}\vbox{\hrule\hbox{\vrule\kern6pt
\parbox{.95\textwidth}{\kern6pt \blue #1 (LP)\vskip6pt}\kern6pt\vrule}\hrule}
        \end{center} \vskip-5mm}

\usepackage[usenames,dvipsnames,svgnames,table]{xcolor}
\usepackage[colorlinks=true,
            linkcolor=red,
            urlcolor=blue,
            citecolor=blue]{hyperref}

\graphicspath{{./}{./Figures/}}

\usepackage{setspace}

\definecolor{lbcolor}{rgb}{0.95,0.95,0.95}
\definecolor{darkred}{RGB}{150,50,50}
\definecolor{brown}{RGB}{250,100,100}
\definecolor{green}{RGB}{000,150,100}
\definecolor{purple}{RGB}{200,000,250}

\def\blue{\color{blue}}

\def\blue{\color{blue}}

\usepackage{indentfirst}

\DeclareMathOperator*{\argmin}{arg\,min}
\DeclareMathOperator*{\argmax}{arg\,max}

\usepackage{color}
\definecolor{red2}{rgb}{0.7, 0, 0.1}

\title{A Distributionally Robust Framework for Safe and Generalizable Utilization of Surrogate Markers}
\date{ }

\author[1,2]{Molei Liu}
\author[3]{Tian Jiang} 
\author[4] {Keyao Zhan\footnote{Liu, Jiang and Zhan had equal contribution to this work.}}
\author[5]{Zirui Zhang}
\author[3,6]{Xuan Wang} 

\affil[1]{Department of Biostatistics, Peking University Health Science Center.}
\affil[2]{Beijing International Center for Mathematical Research, Peking University.}
\affil[3]{Division of Biostatistics, Department of Population Health Sciences, University of Utah.}
\affil[4]{Department of Biostatistics, Harvard Chan School of Public Health.}
\affil[5]{Department of Biostatistics, Columbia University}
\affil[6]{Correspondence: xuan.wang@utah.edu.}

\begin{document}

\maketitle

\begin{abstract}

\noindent Surrogate endpoints can accelerate treatment evaluation, but their validity is challenging to establish because the joint distribution of counterfactual outcome–surrogate pairs is not identifiable. We propose a distributionally robust framework for safe and generalizable surrogate evaluation that explicitly accounts for this uncertainty. By treating cross-arm dependence as partially identified, we formulate a maximin optimization problem that learns an optimal surrogate transformation by maximizing the worst-case explained variance of individual treatment effects over admissible counterfactual dependence structures. The resulting robust surrogacy measure provides a conservative lower bound on explained treatment-effect variation. Moreover, our maximin construction guarantees non-negative alignment between the treatment effects on the transformed surrogate and the primary outcome over the admissible counterfactual couplings. To generalize surrogacy evaluation across populations under covariate shift, we integrate transfer learning with debiased machine learning to estimate the required target-population moments using source trial data and target baseline covariates. The computational tractability and asymptotic properties of our method are established. 
Simulations demonstrate good finite-sample performance and show that the proposed method guards against counterfactual misalignment by shrinking the learned transformation toward zero, revealing a robustness–prediction tradeoff. We illustrate the framework by evaluating 18-month glomerular filtration rate change as a surrogate for two-year change using multiple randomized clinical trial data sets.
 
\vspace{0.3cm}
        \noindent \textbf{Keywords:} Surrogate evaluation; Partial identification; Adversarial optimization; Covariate shift; Debiased machine learning; Counterfactual misalignment.
\end{abstract}

\setstretch{1.5}

\section{Introduction}

\subsection{Background}

Randomized clinical trials remain the gold standard for evaluating new therapies, but definitive clinical endpoints are often difficult to observe in practice. They may require long follow-up, or burdensome measurement, all of which slow the pace of evidence generation. These constraints have motivated the use of surrogate endpoints, namely earlier or easier-to-measure outcomes intended to stand in for the primary clinical endpoint. When a surrogate is reliable, it can shorten trials, reduce cost, and accelerate therapeutic evaluation. Regulatory agencies such as the U.S. Food and Drug Administration (FDA) have established pathways for their use \citep{FDA2014}. Meanwhile, replacing a clinical endpoint with a surrogate is scientifically consequential and requires careful validation.

To determine whether treatment effects on a surrogate adequately capture treatment effects on the primary outcome, one must reason about joint features of potential outcomes across treatment arms. Those joint features are fundamentally unidentifiable because one never observes both treated and untreated outcomes for the same individual. Widely used surrogacy measures may depend on assumptions about cross-world dependence that are not empirically verifiable. The problem is not only technical: a treatment may improve a surrogate while harming the clinical outcome. Surrogate-evaluation procedures should therefore be cautious when the observed data are compatible with counterfactual misalignment, rather than relying on a convenient but untestable point-identifying assumption. A second challenge is generalizability. In practice, surrogate evaluation is often used to inform future studies or to transport evidence from one trial to a target population of interest. Even when a surrogate appears useful in a source study, its performance may change across cohorts with different baseline covariate distributions. Thus, a relevant framework should address both unidentifiable counterfactual dependence and population shift.

\subsection{Related literature}

Statistical surrogate evaluation has been studied under several frameworks. Prentice's operational criterion and proportion-of-treatment-effect-explained measures assess whether adjustment for the surrogate attenuates the treatment effect on the clinical outcome \citep{prentice1989surrogate,freedman1992statistical,wang2002measure}. Meta-analytic methods distinguish individual-level association from trial-level surrogacy \citep{daniels1997,buyse2000validation,burzykowski2005evaluation}. Causal formulations use principal stratification, causal-effect predictiveness, or related criteria \citep{frangakis2002principal,gilbert2008evaluating,joffe2009surrogate}. More recent semiparametric and model-free methods accommodate nonlinear transformations, censoring, and multiple markers \citep{parast2016surrogate,price2018surrogate,wang2020surrogate,wang2021censor,wang2022multi}. Two transformation-based approaches are particularly relevant. \citet{price2018surrogate} define an optimal surrogate by predicting the treatment-specific final outcome, with a primary focus on surrogate construction rather than evaluation. Most closely related, \citet{wang2020surrogate} formulate an oracle mean-square criterion that aims to approximate the individual treatment effect on the outcome by a treatment effect on a transformed surrogate. Because this loss depends on the unobserved joint distribution of the two counterfactual outcome--surrogate pairs, they derive a tractable transformation under a working cross-world independence condition. Our method instead retains the original same-individual counterfactual loss and optimizes it over a structured class of admissible cross-arm dependence. 

Our formulation is closely related to partial identification because randomization identifies the treatment-specific potential-outcome distributions, but not their joint cross-arm coupling. Consequently, features of individual treatment effects that depend on this coupling are generally only partially identified \citep{manski2003,par2010tamer,par2023kline,aronow2014,firpo2019partial}. Classical work characterizes such identified sets and develops inference for their bounds \citep{imbens2004partial}, while more recent methods use covariates, optimal transport, and flexible nuisance estimation to sharpen partially identified causal analyses \citep{ji2023causal,jiang2025causal}. Our problem differs in that the target itself is optimized: rather than estimating bounds for a prespecified causal functional, we jointly choose the surrogate transformation and the least favorable cross-arm dependence structure, yielding a maximin surrogate utility. 

Distributionally robust optimization provides a natural framework for decision-making under partial identification by optimizing performance over an ambiguity set of distributions \citep{scarf1958dro,rahimian2022dro}. In causal inference, ambiguity sets have been used to represent unobserved counterfactual distributions or departures from identifying assumptions. \citet{guo2022noisy}, for example, formulate partial identification with noisy covariates as robust optimization over latent data-generating distributions. Related work uses distributional uncertainty sets for sensitivity analysis under unmeasured confounding \citep{bertsimas2023drocausal,jin2026fsensitivity,dorn2025dvds}, to tighten bounds over admissible potential-outcome couplings \citep{lin2025covariateOT}, and to construct robust causal targets when standard identifying assumptions may fail \citep{koo2025synthetic}. These papers share our high-level strategy of propagating ambiguity about an unobserved causal distribution through an optimization problem rather than selecting one convenient completion. Our construction differs in the source of ambiguity and the decision variable. The uncertainty in our setting is concentrated on the cross-world dependence that remains unidentified even under randomization. 

Another related field is generalizable surrogacy evaluation across populations. Meta-analytic validation evaluates surrogacy across completed trials, while recent resilience analyses allow the surrogate--outcome relationship in a future study to deviate from that learned previously \citep{buyse2000validation,inker2023meta}. Transformation-based work also studies conditions under which an optimal surrogate learned in one trial remains valid in another \citep{price2018surrogate}. Our setting is different but complementary: a source randomized trial contains the outcome and surrogate, whereas the target sample may contain only baseline covariates, and the conditional outcome model is transported under covariate shift. 

Moreover, existing literature studies the surrogate paradox, in which favorable behavior of the surrogate may fail to correspond to a favorable treatment effect on the primary outcome \citep{elliott2023review}. \citet{chen2007paradox} derived criteria under which a favorable treatment effect on the surrogate can coexist with an unfavorable effect on the primary outcome. \citet{vanderweele2013surrogate} showed that positive surrogate--outcome association does not generally guarantee preservation of the direction of treatment effects. \citet{yin2020criteria} subsequently obtained testable conditions that are sufficient and almost necessary, relative to the observed-data distribution, for excluding the classical average-effect paradox. \citet{ma2021individual} studied the individual surrogate paradox and derived sharp bounds for the proportion of individuals for whom favorable surrogate behavior can mask harm on the primary outcome. \citet{hsiao2026avoid} developed empirically assessable sufficient conditions for avoiding the surrogate paradox, while \citet{hsiao2025resilience} studied the resilience of the surrogate paradox under controlled departures from cross-study assumptions. Our counterfactual-alignment criterion is complementary to these developments. It assesses whether the learned surrogate transformation remains positively aligned with the individual treatment effect on the primary outcome over the specified ambiguity class.

\subsection{Our contributions}

We propose a distributionally robust framework for surrogate evaluation that is designed to be safe and generalizable. We begin with a causal surrogacy criterion based on the explained variance of the individual treatment effect by an optimal transformation of the surrogate. Because the joint distribution of counterfactual outcome--surrogate pairs across the treatment and control arms is not identifiable, we do not impose a single working model for cross-arm dependence. Instead, we define an ambiguity set of dependence structures consistent with the observed within-arm information and choose the surrogate transformation that maximizes the worst-case explained variance over that set. This yields a robust surrogacy measure anchored to the identifiable content of the data. We also develop a transfer learning version of the framework for a target population with a different covariate distribution. Under covariate shift, we combine density-ratio estimation with conditional moment modeling to construct debiased, cross-fitted estimators of the target-population moments entering the robust objective. This allows the proposed surrogacy measure to be learned from a source randomized trial and transported to a target population using flexible machine learning.

Our proposed DRO formulation has an important practical implication: it naturally guards against unsafe surrogate use. When the observed data are compatible with counterfactual misalignment, the robust explained-variance criterion shrinks toward zero, reflecting limited support for using the surrogate to approximate the treatment effect on the primary outcome. For computation, we show that the resulting maximin problem admits a convenient semidefinite-program formulation for the operator-norm ambiguity set used in our implementation. To stabilize estimation and inference when the least favorable dependence structure may not be unique, we further introduce a ridge-regularized version of the criterion.

On the theoretical side, we characterize the population maximin solution, establish consistency of the proposed estimators, and develop asymptotically valid inference for the ridge-stabilized robust surrogate utility and the optimal transformation. On the empirical side, simulation studies show that the method has good finite-sample performance for estimation and inference, shrinks the learned transformation toward the null when the observed data are compatible with counterfactual misalignment, and exhibits a robustness--prediction tradeoff. We illustrate the proposed method by evaluating 18-month glomerular filtration rate (GFR) change as a surrogate for two-year GFR change using randomized clinical trial data from chronic kidney disease studies. Together, these developments provide a principled way to evaluate and transport surrogacy while explicitly accounting for uncertainty in unobserved counterfactual dependence.

\section{Target estimand and partial identification}\label{sec:notation}

\subsection{Study setup and surrogate utility}

Let $A\in\{0,1\}$ denote treatment, $Y$ the primary outcome, $S$ the candidate surrogates, and $\boldsymbol X\in\mathbb R^p$ baseline covariates. For $a\in\{0,1\}$, let $(Y^{(a)},S^{(a)})$ be the potential outcome--surrogate pair under treatment level $a$. From the source randomized trial, we observe independent copies $O_i=(\boldsymbol X_i,A_i,Y_i,S_i)$, $i=1,\ldots,n$, where $Y_i=A_iY_i^{(1)}+(1-A_i)Y_i^{(0)}$ and $S_i=A_iS_i^{(1)}+(1-A_i)S_i^{(0)}$. Write $\mathbb P$ for the source distribution and $\mathbb P_X$ for its covariate marginal. The randomization of the trial gives that 
\[
A\perp\!\!\!\perp\{Y^{(1)},Y^{(0)},S^{(1)},S^{(0)},\boldsymbol X\}\quad\text{under }\mathbb P.
\]
Suppose that we additionally observe target covariates $\boldsymbol X^{\mathrm{tar}}_1,\ldots,\boldsymbol X^{\mathrm{tar}}_N\overset{\mathrm{iid}}{\sim}\mathbb Q_X$, where $\mathbb Q_X$ may differ from $\mathbb P_X$. We assume that the conditional distribution of the potential outcomes (outcome--surrogate pairs) given $\boldsymbol X$ is common to the source and target populations, so that only the marginal covariate distribution changes.

Within this target population, our goal is to derive a transformation of the surrogate that approximates the individual treatment effect on the primary outcome. Let $\boldsymbol\phi(s)\in\mathbb R^d$ be a prespecified basis and write $g_{\boldsymbol\beta}(s)=\boldsymbol\phi(s)^\top\boldsymbol\beta$ for $\boldsymbol\beta\in\mathbb R^d$, where $\boldsymbol\phi(s)=s$ is a special case. Define
$\boldsymbol Z^{(a)}=\{Y^{(a)},\boldsymbol\phi(S^{(a)})^\top\}^\top$, $\Delta_Y=Y^{(1)}-Y^{(0)}$, $\boldsymbol\Delta_\phi=\boldsymbol\phi(S^{(1)})-\boldsymbol\phi(S^{(0)})$, and $\Delta_{g_{\boldsymbol\beta}}=\Delta_{\boldsymbol\beta}=\boldsymbol\beta^\top\boldsymbol\Delta_\phi$. For a candidate full-data distribution $\mathbb Q$, define the surrogate utility as
\begin{equation}
\mathcal R(\boldsymbol\beta;\mathbb Q)
=\mathbb E_{\mathbb Q}(\Delta_Y^2)-\mathbb E_{\mathbb Q}\{(\Delta_Y-\boldsymbol\beta^\top\boldsymbol\Delta_\phi)^2\}
=2\boldsymbol\beta^\top\mathbb E_{\mathbb Q}(\boldsymbol\Delta_\phi\Delta_Y)
-\boldsymbol\beta^\top\mathbb E_{\mathbb Q}(\boldsymbol\Delta_\phi\boldsymbol\Delta_\phi^\top)\boldsymbol\beta.
\label{eq:utility}
\end{equation}
Here, $\mathcal R(\boldsymbol\beta;\mathbb Q)$ is the reduction in mean-squared error achieved by using the transformed surrogate to approximate the individual treatment effect. The normalized version is defined as the surrogacy measure $\tau(\boldsymbol\beta;\mathbb Q)=\mathcal R(\boldsymbol\beta;\mathbb Q)/\mathbb E_{\mathbb Q}(\Delta_Y^2)$ when the denominator is positive. We use $\mathcal R$ as the primitive criterion because it isolates the key identification issue without introducing an additional unidentified denominator. The null transformation $\boldsymbol\beta=\boldsymbol0$ corresponds to ignoring the surrogate and has utility zero. Positive utility therefore represents a genuine improvement over this safety benchmark, whereas negative utility means that the transformed surrogate is more misleading than using no surrogate at all. This mean-square formulation is close to that of \citet{wang2020surrogate}, but the treatment of nonidentifiability is fundamentally different. We do not impose working independence between the two counterfactual outcome--surrogate pairs. The unidentified cross-arm dependence remains inside the original loss and is handled explicitly by the ambiguity set introduced below.

\subsection{Partial identification through counterfactual dependence}\label{sec:frame:partial}

The source trial identifies the arm-specific conditional moments
$
\boldsymbol m_a(\boldsymbol x)=\mathbb E_{\mathbb P}(\boldsymbol Z^{(a)}\mid\boldsymbol X=\boldsymbol x),
$
and
$
\boldsymbol H_a(\boldsymbol x)=\mathbb E_{\mathbb P}(\boldsymbol Z^{(a)}\boldsymbol Z^{(a)\top}\mid\boldsymbol X=\boldsymbol x),
$
with conditional covariance $\boldsymbol V_a(\boldsymbol x)=\boldsymbol H_a(\boldsymbol x)-\boldsymbol m_a(\boldsymbol x)\boldsymbol m_a(\boldsymbol x)^\top$. Averaging these quantities over the target covariate distribution gives the identified target within-arm moments
\begin{equation}
\boldsymbol\mu_a=\mathbb E_{\mathbb Q_X}\{\boldsymbol m_a(\boldsymbol X)\},\qquad
\boldsymbol\Sigma_a=\mathbb E_{\mathbb Q_X}\{\boldsymbol H_a(\boldsymbol X)\}-\boldsymbol\mu_a\boldsymbol\mu_a^\top,
\qquad a\in\{0,1\}.
\label{eq:within-arm-moments}
\end{equation}
These arm-specific quantities are identifiable, but the cross-arm covariance is not because $(\boldsymbol Z^{(1)},\boldsymbol Z^{(0)})$ is never jointly observed. This unidentified conditional dependence structure is
\[
\operatorname{Cov}\{\boldsymbol Z^{(1)},\boldsymbol Z^{(0)}\mid\boldsymbol X=\boldsymbol x\}
=\boldsymbol V_1(\boldsymbol x)^{1/2}\boldsymbol\Gamma(\boldsymbol x)\boldsymbol V_0(\boldsymbol x)^{1/2},
\]
where $\boldsymbol\Gamma(\boldsymbol x) \in \mathbb{R}^{(1+d)\times(1+d)}$ satifies that $\|\boldsymbol\Gamma(\boldsymbol x)\|_{\mathrm{op}}\le1$. Here, $\boldsymbol\Gamma(\boldsymbol x)$ denotes an unidentified matrix that characterizes the residual correlation between $\boldsymbol Z^{(1)}$ and $\boldsymbol Z^{(0)}$ conditional on $\boldsymbol X=\boldsymbol x$. The operator-norm constraint $\|\boldsymbol\Gamma(\boldsymbol x)\|_{\mathrm{op}}\le 1$ ensures that the resulting joint conditional covariance matrix of $(\boldsymbol Z^{(1)},\boldsymbol Z^{(0)})$ is positive semidefinite and hence corresponds to a valid dependence structure. Consequently, the target cross-arm covariance is
\begin{equation}
\boldsymbol\Sigma_{10}
=\mathbb E_{\mathbb Q_X}\{\boldsymbol V_1(\boldsymbol X)^{1/2}\boldsymbol\Gamma(\boldsymbol X)\boldsymbol V_0(\boldsymbol X)^{1/2}\}
+\operatorname{Cov}_{\mathbb Q_X}\{\boldsymbol m_1(\boldsymbol X),\boldsymbol m_0(\boldsymbol X)\}.
\label{equ:cross:cov}
\end{equation}
The second term is identified through covariate-dependent conditional means, whereas the first captures the residual ambiguity in counterfactual dependence. This decomposition clarifies the role of baseline covariates. Part of the cross-arm association is induced by the common dependence of the two potential outcome--surrogate vectors on baseline covariates and is therefore identified through the conditional means $\boldsymbol m_1(\boldsymbol X)$ and $\boldsymbol m_0(\boldsymbol X)$. The remaining association, after conditioning on $\boldsymbol X$, is represented by the residual dependence structure and remains unidentified. In this way, baseline covariates contribute to separating identified from unidentified components of the cross-arm dependence.

Let $\mathcal Q$ collect all full-data distributions that agree with the identified arm-specific conditional distributions and the target covariate distribution $\mathbb Q_X$. For a fixed $\boldsymbol\beta$, the observed data therefore determine the set $\{\mathcal R(\boldsymbol\beta;\mathbb Q):\mathbb Q\in\mathcal Q\}$ rather than a single value. This partial-identification perspective motivates the robust criterion developed next. A conventional point-identifying analysis selects one element of this set, typically by imposing a working model on the unobserved coupling. Our approach instead treats the variation across compatible couplings as genuine scientific uncertainty. The next section investigates what surrogate transformation can still be justified without choosing a favorable, unverifiable cross-world distribution from $\mathcal Q$.

\section{Distributionally robust framework}

\subsection{Robust surrogate evaluation}\label{sec:robust-evaluation}

For computational tractability, we restrict the covariate-dependent coupling matrix $\boldsymbol\Gamma(\boldsymbol X)$ introduced in Section \ref{sec:frame:partial} to be constant across $\boldsymbol X$, i.e., $\Gamma(\boldsymbol X) \equiv \boldsymbol\Gamma$. Although we adopt this simplification throughout, the proposed framework can in principle be extended to accommodate
$\boldsymbol X$-dependent coupling structures at the cost of increased computational and theoretical complexity. Specifically, we assume $\boldsymbol\Gamma(\boldsymbol x)\equiv\boldsymbol\Gamma\in\mathcal G$, where $\mathcal G$ is a compact subset of $\{\boldsymbol\Gamma\in\mathbb R^{(1+d)\times(1+d)}:\|\boldsymbol\Gamma\|_{\mathrm{op}}\le1-\varepsilon_\Gamma\}$ for some $\varepsilon_\Gamma>0$. 
The set $\mathcal G$ denotes which cross-arm dependence structures the analyst is willing to regard as scientifically plausible. The full operator-norm ball provides a conservative default, while narrower choices can encode credible structural information. 
Let $\mathcal Q_{\mathcal G}\subset\mathcal Q$ denote the class of target full-data distributions whose conditional cross-arm covariance is represented by some $\boldsymbol\Gamma\in\mathcal G$. All robustness statements below are with respect to $\mathcal Q_{\mathcal G}$. Under this restriction, we can write
\begin{equation}
\boldsymbol\Sigma_{10}(\boldsymbol\Gamma)
=\mathbb E_{\mathbb Q_X}\{\boldsymbol V_1(\boldsymbol X)^{1/2}\boldsymbol\Gamma\boldsymbol V_0(\boldsymbol X)^{1/2}\}
+\operatorname{Cov}_{\mathbb Q_X}\{\boldsymbol m_1(\boldsymbol X),\boldsymbol m_0(\boldsymbol X)\}.
\label{cross}
\end{equation}
Then, the second moment of $(\Delta_Y,\boldsymbol\Delta_\phi^\top)^\top$ is expressed as
\[
\boldsymbol M(\boldsymbol\Gamma)=\boldsymbol\Sigma_1+\boldsymbol\Sigma_0-\boldsymbol\Sigma_{10}(\boldsymbol\Gamma)-\boldsymbol\Sigma_{10}(\boldsymbol\Gamma)^\top
+(\boldsymbol\mu_1-\boldsymbol\mu_0)(\boldsymbol\mu_1-\boldsymbol\mu_0)^\top.
\]


Let $\boldsymbol c(\boldsymbol\Gamma)$ and $\boldsymbol G(\boldsymbol\Gamma)$ denote the lower-left and lower-right blocks of $\boldsymbol M(\boldsymbol\Gamma)$, respectively. 
The surrogate surrogate utility function under a given coupling structure $\boldsymbol c(\boldsymbol\Gamma)$ and $\boldsymbol G(\boldsymbol\Gamma)$ can be equivalently written as 
\[
\mathcal R(\boldsymbol\beta,\boldsymbol\Gamma)
=2\boldsymbol\beta^\top\boldsymbol c(\boldsymbol\Gamma)-\boldsymbol\beta^\top\boldsymbol G(\boldsymbol\Gamma)\boldsymbol\beta,
\]
and the corresponding maximin problem is
\begin{equation}
\boldsymbol\beta^*\in\argmax_{\boldsymbol\beta\in\mathbb R^d}
\min_{\boldsymbol\Gamma\in\mathcal G}\mathcal R(\boldsymbol\beta,\boldsymbol\Gamma),
\quad
\mathcal R^*=\min_{\boldsymbol\Gamma\in\mathcal G}\mathcal R(\boldsymbol\beta^*,\boldsymbol\Gamma).
\label{equ:maximin}
\end{equation}
The maximin criterion selects a single surrogate transformation by maximizing its worst-case utility over all dependence structures in $\mathcal G$. Thus, $\mathcal R^*$ quantifies the utility that can be guaranteed without committing to a particular cross-arm coupling. Since $\boldsymbol\beta=\boldsymbol0$ is always feasible and yields zero utility, $\mathcal R^*\ge0$; values near zero indicate that the observed data provide little robust support for a nontrivial surrogate transformation.

\begin{assumption}
\label{assump:pd-var}
For $a\in\{0,1\}$, $c_V I \preceq \boldsymbol V_a(x) \preceq C_V I$ almost surely. The set $\mathcal G$ is compact and convex and satisfies $\|\boldsymbol\Gamma\|_{\mathrm{op}}\le1-\varepsilon_\Gamma$ for every $\boldsymbol\Gamma\in\mathcal G$.
\end{assumption}

The two parts of Assumption~\ref{assump:pd-var} are regularity conditions on the within-arm variation and the ambiguity set. Uniform positive definiteness of $\boldsymbol V_a(\boldsymbol x)$ rules out degenerate directions in the outcome--surrogate vector, while the strict operator-norm margin keeps admissible cross-arm couplings away from deterministic dependence and ensures a nonsingular joint conditional covariance structure. Compactness of $\mathcal G$ guarantees existence of a least-favorable coupling, and convexity supports the minimax characterization below. None of these conditions favors positive alignment between the surrogate and the outcome.

For a fixed $\boldsymbol\Gamma$, maximizing over $\boldsymbol\beta$ gives $\boldsymbol\beta(\boldsymbol\Gamma)=\boldsymbol G(\boldsymbol\Gamma)^{-1}\boldsymbol c(\boldsymbol\Gamma)$ and the profiled objective
\begin{equation}
q(\boldsymbol\Gamma)=\boldsymbol c(\boldsymbol\Gamma)^\top\boldsymbol G(\boldsymbol\Gamma)^{-1}\boldsymbol c(\boldsymbol\Gamma).
\label{eq:mingamma}
\end{equation}

\begin{theorem}[Population solution]
\label{thm:popu-solution}
Under Assumption~\ref{assump:pd-var}, $\boldsymbol G(\boldsymbol\Gamma)$ is positive definite for every $\boldsymbol\Gamma\in\mathcal G$. Moreover,
\begin{equation}
\boldsymbol\Gamma^*\in\argmin_{\boldsymbol\Gamma\in\mathcal G}q(\boldsymbol\Gamma),\qquad
\boldsymbol\beta^*=\boldsymbol G(\boldsymbol\Gamma^*)^{-1}\boldsymbol c(\boldsymbol\Gamma^*),\qquad
\mathcal R^*=q(\boldsymbol\Gamma^*).
\label{eq:betastar}
\end{equation}
The transformation $\boldsymbol\beta^*$ and the value $\mathcal R^*$ are unique, although the least-favorable $\boldsymbol\Gamma^*$ need not be.
\end{theorem}
The theorem makes the robust construction transparent. For each admissible coupling, $q(\boldsymbol\Gamma)$ is the maximum surrogate utility attainable under that coupling. The proposed method selects the coupling for which this optimal utility is smallest and uses the corresponding transformation. It therefore avoids committing to a favorable working model for the unobserved cross-arm dependence. The distributional uncertainty here is confined to the unidentified part of the full-data distribution: all admissible distributions agree on the treatment-arm-specific distributions identified by the randomized trial and differ only in their cross-arm coupling. Working independence is therefore one admissible coupling rather than a point-identifying assumption, and the original same-individual counterfactual loss is retained throughout. The procedure jointly learns the surrogate transformation and evaluates its worst-case utility over the admissible couplings, rather than merely stress-testing a prespecified transformation.

For a unit-free summary, let $D(\mathbb Q)=\mathbb E_{\mathbb Q}(\Delta_Y^2)$. Since $D(\mathbb Q)$ is also partially identified, we use the identified upper bound
\[
\overline D=\mathbb E_{\mathbb Q_X}\!\left[\{m_{Y,1}(\boldsymbol X)-m_{Y,0}(\boldsymbol X)\}^2+
\{\sigma_{Y,1}(\boldsymbol X)+\sigma_{Y,0}(\boldsymbol X)\}^2\right],
\]
where $m_{Y,a}(\boldsymbol x)=\mathbb E_{\mathbb P}(Y^{(a)}\mid\boldsymbol X=\boldsymbol x)$ and $\sigma^2_{Y,a}(\boldsymbol x)=\operatorname{Var}_{\mathbb P}(Y^{(a)}\mid\boldsymbol X=\boldsymbol x)$. We report $\tau^*=\mathcal R^*/\overline D$ as a normalized surrogacy measure and also report the unnormalized utility $\mathcal R^*$, which facilitates comparison with existing methods defined on the original scale. The normalization improves comparability across outcome scales, but it is intentionally conservative because $\overline D$ is an upper bound on the unidentified second moment $\mathbb E_{\mathbb Q}(\Delta_Y^2)$. Thus, $\tau^*$ provides a unit-free summary of surrogacy, while $\mathcal R^*$ retains the original-scale criterion for comparison with related approaches.

Although the robust surrogate utility and transformation are well defined, the least-favorable coupling need not be unique. This can destabilize numerical optimization and create nonregularity in inference. We therefore introduce a ridge penalty to select a unique, stabilized least-favorable coupling. For $\lambda>0$, define
\begin{equation}
\boldsymbol\Gamma_\lambda^*=\argmin_{\boldsymbol\Gamma\in\mathcal G}\{q(\boldsymbol\Gamma)+\lambda\|\boldsymbol\Gamma\|_F^2\},
\qquad
\boldsymbol\beta_\lambda^*=\boldsymbol G(\boldsymbol\Gamma_\lambda^*)^{-1}\boldsymbol c(\boldsymbol\Gamma_\lambda^*),
\label{ridge}
\end{equation}
and let $\mathcal R_\lambda^*
=q(\boldsymbol\Gamma_\lambda^*)
+\lambda\|\boldsymbol\Gamma_\lambda^*\|_F^2$
denote the ridge-stabilized robust surrogate utility used for inference. Its normalized counterpart is
\(
\tau_\lambda^*=\mathcal R_\lambda^*/\overline D,
\)
which we call the ridge-stabilized robust surrogacy measure. The unpenalized value $q(\boldsymbol\Gamma_\lambda^*)$ remains the surrogate utility at the stabilized solution. The ridge term is introduced to regularize the least-favorable coupling, not to assert that small cross-arm dependence is scientifically more plausible. It stabilizes the optimizer when the unpenalized problem has multiple or nearly equivalent least-favorable couplings and yields a regularized target for inference.

\begin{proposition}[Ridge approximation error]
\label{prop:ridge control}
For every $\lambda>0$,
\begin{equation}
0\le q(\boldsymbol\Gamma_\lambda^*)-\mathcal R^*
\le\lambda\min_{\boldsymbol\Gamma^*\in\argmin_{\boldsymbol\Gamma\in\mathcal G}q(\boldsymbol\Gamma)}\|\boldsymbol\Gamma^*\|_F^2
\le\lambda(1+d).
\label{eq:comparison}
\end{equation}
\end{proposition}

Thus, ridge stabilization changes the surrogate utility evaluated at the stabilized coupling by at most $O(\lambda)$ while stabilizing the least-favorable coupling.

\subsection{Counterfactual alignment and paradox protection}

The same robust criterion also provides a direct safety interpretation. Indeed,
\begin{equation}
\mathcal R(\boldsymbol\beta;\mathbb Q)
=2\mathbb E_{\mathbb Q}(\Delta_{\boldsymbol\beta}\Delta_Y)
-\mathbb E_{\mathbb Q}(\Delta_{\boldsymbol\beta}^2).
\label{eq:risk-alignment}
\end{equation}
Thus, rather than evaluating alignment under one unverifiable counterfactual coupling, the robust criterion evaluates whether the treatment effect on the transformed surrogate remains positively aligned, in expectation, with the treatment effect on the primary outcome over the admissible counterfactual couplings.

\begin{definition}[Counterfactual misalignment]
\label{def:alignment-paradox}
For a fixed $\mathbb Q\in\mathcal Q_{\mathcal G}$, $g_{\boldsymbol\beta}$ exhibits counterfactual misalignment under $\mathbb Q$ if
$\mathbb E_{\mathbb Q}(\Delta_{\boldsymbol\beta}\Delta_Y)\le0$.
It is compatible with counterfactual misalignment over $\mathcal Q_{\mathcal G}$ if this inequality holds for at least one compatible distribution.
\end{definition}

\begin{proposition}[Protection against counterfactual misalignment]
\label{prop:paradox-protection}
If
$
\inf_{\mathbb Q\in\mathcal Q_{\mathcal G}}
\mathcal R(\boldsymbol\beta;\mathbb Q)>0,
$
then we have
$
\inf_{\mathbb Q\in\mathcal Q_{\mathcal G}}
\mathbb E_{\mathbb Q}(\Delta_{\boldsymbol\beta}\Delta_Y)>0.
$
Conversely, if $g_{\boldsymbol\beta}$ is compatible with counterfactual misalignment over $\mathcal Q_{\mathcal G}$, then
\[
\inf_{\mathbb Q\in\mathcal Q_{\mathcal G}}
\mathcal R(\boldsymbol\beta;\mathbb Q)\le0.
\]
\end{proposition}

The result follows immediately from \eqref{eq:risk-alignment}. Since the null transformation has worst-case utility zero, the maximin rule can shrink toward $\boldsymbol\beta=\boldsymbol0$ rather than endorse a transformation whose apparent usefulness depends on a favorable cross-world assumption. 
The protection is deliberately one-sided. A positive worst-case surrogate utility certifies that positive alignment holds uniformly over the compatible couplings, but a nonpositive value does not prove that the surrogate is invalid under the unknown true coupling. It means that the observed data and maintained restrictions are insufficient to certify positive counterfactual alignment.

The classical mean-effect paradox concerns the signs of $\mathbb E_{\mathbb Q}(\Delta_{\boldsymbol\beta})$ and $\mathbb E_{\mathbb Q}(\Delta_Y)$ \citep{vanderweele2013surrogate}, both of which are identified here. Our criterion instead addresses the treatment-contrast-specific question of whether the individual treatment effects on the transformed surrogate and primary outcome are positively aligned in expectation.
The preceding result concerns counterfactual alignment and does not by itself exclude the conventional mean-effect surrogate paradox \citep{vanderweele2013surrogate}. For completeness, Appendix~\ref{app:classical-paradox} gives identified sufficient conditions under which the learned transformation also avoids the conventional surrogate paradox.

\section{Debiased estimation under covariate shift}

The robust criterion depends on target moments transported from the source trial. A direct plug-in estimator can inherit first-order bias from flexible nuisance estimation, especially through the nonlinear cross-arm covariance. We therefore construct cross-fitted one-step estimators of the required moments and substitute them into the same profiled optimization used at the population level. The estimation strategy has three modular steps. We first estimate arm-specific conditional moments from the randomized trial and average them over the target covariate distribution. We then orthogonalize these transported moments so that errors from flexible nuisance estimation enter only at second order. Finally, we solve the same low-dimensional robust optimization as at the population level. This separation clarifies the distinct roles of estimation and robustness: machine learning is used to recover identifiable conditional moments, while robustness addresses the distinct problem of unidentified counterfactual dependence.

\subsection{Debiased transport of target moments}\label{sec:debiased-moments}

Let $\pi_a=\mathbb P(A=a)$, $\xi_a(A)=\mathbbm 1(A=a)/\pi_a$, and $w(\boldsymbol x)=d\mathbb Q_X(\boldsymbol x)/d\mathbb P_X(\boldsymbol x)$. For functions $f$ and $h$, write
$
\mathbb P_n f=n^{-1}\sum_{i=1}^n f(O_i),
$ 
and 
$
\mathbb Q_N h=N^{-1}\sum_{j=1}^N h(\boldsymbol X_j^{\mathrm{tar}}).
$
All nuisance estimators below are cross-fitted, with fold indices suppressed for readability. We first estimate the identified within-arm moments. Let $\widehat w$, $\widehat{\boldsymbol m}_a$, and $\widehat{\boldsymbol H}_a$ estimate $w$, $\boldsymbol m_a$, and $\boldsymbol H_a$. The augmented transport estimators are
\begin{align}
\widehat{\boldsymbol\mu}_a
&=\mathbb Q_N\{\widehat{\boldsymbol m}_a(\boldsymbol X)\}
+\mathbb P_n\!\left[\xi_a(A)\widehat w(\boldsymbol X)\{\boldsymbol Z-\widehat{\boldsymbol m}_a(\boldsymbol X)\}\right],
\label{eq:muhat}\\[-2pt]
\widehat{\boldsymbol\Sigma}_a
&=\mathbb Q_N\{\widehat{\boldsymbol H}_a(\boldsymbol X)\}
+\mathbb P_n\!\left[\xi_a(A)\widehat w(\boldsymbol X)\{\boldsymbol Z\boldsymbol Z^\top-\widehat{\boldsymbol H}_a(\boldsymbol X)\}\right]
-\widehat{\boldsymbol\mu}_a\widehat{\boldsymbol\mu}_a^\top.
\label{eq:Omegahat}
\end{align}
The target-average term transports the fitted conditional moment, while the residual term provides a first-order bias correction. Because treatment is randomized, no propensity-score model is required. The augmentation combines two representations of the same target moment: imputation over the target covariates and a density-ratio-weighted residual correction from the source trial. The resulting orthogonal score makes the estimator insensitive to first-order nuisance estimation error, allowing the conditional moments and density ratio to be estimated flexibly. Randomization removes the need to estimate treatment assignment, while the source-to-target density ratio accounts for the covariate shift between the two populations.

The cross-arm covariance in \eqref{cross} is a smooth but nonlinear functional of $w$, $\boldsymbol m_a$, and $\boldsymbol V_a$. With $\widehat{\boldsymbol V}_a=\widehat{\boldsymbol H}_a-\widehat{\boldsymbol m}_a\widehat{\boldsymbol m}_a^\top$, we estimate it by
\begin{equation}
\widehat{\boldsymbol\Sigma}_{10}(\boldsymbol\Gamma)
=\mathbb Q_N\!\left[\widehat{\boldsymbol V}_1(\boldsymbol X)^{1/2}\boldsymbol\Gamma\widehat{\boldsymbol V}_0(\boldsymbol X)^{1/2}
+\widehat{\boldsymbol m}_1(\boldsymbol X)\widehat{\boldsymbol m}_0(\boldsymbol X)^\top\right]
-\widehat{\boldsymbol\mu}_1\widehat{\boldsymbol\mu}_0^\top
+\mathbb P_n\{\widehat{\boldsymbol\psi}_{10}(O;\boldsymbol\Gamma)\}.
\label{eq:Sigma10hat}
\end{equation}
The one-step correction $\widehat{\boldsymbol\psi}_{10}$ accounts for estimation of the conditional means, density ratio, and matrix square roots. Its explicit form follows from the Fr\'echet derivative of the matrix square-root map and is given in Appendix~\ref{app:cross-cov-estimator}. This cross-arm term is the main technical departure from standard transported-mean estimation. The robust objective depends on covariance square roots over the entire ambiguity set, so plug-in estimation error propagates through a nonlinear matrix map and can affect the estimated worst-case coupling at first order. The one-step correction is therefore designed to estimate the entire moment surface indexed by $\boldsymbol\Gamma$ with first-order accuracy, rather than correcting the covariance at only a fixed value of $\boldsymbol\Gamma$.

\subsection{Empirical optimization and implementation}\label{sec:crossfit}

Having estimated the target moments, we now form the empirical analogue of the population optimization by substituting the preceding debiased estimators into $\boldsymbol M(\boldsymbol\Gamma)$, yielding $\widehat{\boldsymbol M}(\boldsymbol\Gamma)$; let $\widehat{\boldsymbol c}(\boldsymbol\Gamma)$ and $\widehat{\boldsymbol G}(\boldsymbol\Gamma)$ denote its lower-left and lower-right blocks.

Define
\begin{equation}
\widehat q(\boldsymbol\Gamma)=\widehat{\boldsymbol c}(\boldsymbol\Gamma)^\top\widehat{\boldsymbol G}(\boldsymbol\Gamma)^{-1}\widehat{\boldsymbol c}(\boldsymbol\Gamma),\qquad
\widehat{\boldsymbol\Gamma}\in\argmin_{\boldsymbol\Gamma\in\mathcal G}\widehat q(\boldsymbol\Gamma),\qquad
\widehat{\boldsymbol\beta}=\widehat{\boldsymbol G}(\widehat{\boldsymbol\Gamma})^{-1}\widehat{\boldsymbol c}(\widehat{\boldsymbol\Gamma}).
\label{qhat}
\end{equation}
The unpenalized estimator of the robust surrogate utility is
$\widehat{\mathcal R}=\widehat q(\widehat{\boldsymbol\Gamma})$. For the stabilized estimator,
\begin{equation}
\widehat{\boldsymbol\Gamma}_\lambda
=\argmin_{\boldsymbol\Gamma\in\mathcal G}\{\widehat q(\boldsymbol\Gamma)+\lambda\|\boldsymbol\Gamma\|_F^2\},\qquad
\widehat{\boldsymbol\beta}_\lambda
=\widehat{\boldsymbol G}(\widehat{\boldsymbol\Gamma}_\lambda)^{-1}\widehat{\boldsymbol c}(\widehat{\boldsymbol\Gamma}_\lambda),
\label{eq:ridge-estimator}
\end{equation}
with penalized value $\widehat{\mathcal R}_\lambda=\widehat q(\widehat{\boldsymbol\Gamma}_\lambda)+\lambda\|\widehat{\boldsymbol\Gamma}_\lambda\|_F^2$. Here we estimate the normalizing quantity $\overline D$ using the corresponding cross-fitted one-step estimator described and denote it by $\widehat{\overline D}$. In applications, we report both the ridge-stabilized robust surrogate utility $\widehat{\mathcal R}_\lambda$ and its normalized version $\widehat\tau_\lambda=\widehat{\mathcal R}_\lambda/\widehat{\overline D}$.

For cross-fitting, we split both samples into $K$ folds, estimate the nuisance functions on the complementary folds, evaluate the one-step scores on the held-out folds, and average the resulting moment estimates. This construction permits generic machine-learning estimators without Donsker conditions and renders nuisance-estimation error second order under the rate conditions imposed in Section~\ref{sec:theory}.

\section{Theoretical properties}\label{sec:theory}

Throughout, $n,N\to\infty$ with $N/n\to\kappa\in(0,\infty)$, and $\lambda>0$ is fixed unless stated otherwise. The results below separate three inferential goals. Consistency is available for both the original maximin surrogate utility and its ridge-stabilized counterpart. Regular inference is first developed for the ridge-stabilized robust surrogate utility, which remains well behaved even when the least-favorable coupling is constrained. Inference for the transformation itself is more demanding because it must track the optimizer and therefore requires an additional interiority condition.

\begin{assumption}[Moments and overlap]\label{assump:bound}
For some $\nu>0$, $\mathbb E_{\mathbb P}\|\boldsymbol Z\|^{4+\nu}<\infty$. The density ratio $w=d\mathbb Q_X/d\mathbb P_X$ exists and is uniformly bounded.
\end{assumption}

The overlap requirement concerns generalization from the source population to the target population, not treatment positivity within the randomized trial. Bounded $w$ requires target covariate profiles to be represented in the source trial and rules out unrestricted extrapolation beyond its support. When the density ratio is highly variable, transport remains identified but becomes statistically less stable because density-ratio weighting increases variability in the source-sample contribution to the estimator.


\begin{assumption}[Nuisance estimation]\label{assump:nuisance}
For some $a_n\to0$, the cross-fitted estimators introduced in Section~4 satisfy
\[
\max_{a\in\{0,1\}}
\left\{
\|\widehat w-w\|_{L_2},
\|\widehat{\boldsymbol m}_a-\boldsymbol m_a\|_{L_4},
\|\widehat{\boldsymbol H}_a-\boldsymbol H_a\|_{L_2},
\|\widehat{\boldsymbol V}_a-\boldsymbol V_a\|_{L_4}
\right\}
=O_p(a_n),\qquad \sqrt n\,a_n^2\to0.
\]
In addition, the estimated conditional covariance matrices $\widehat{\boldsymbol V}_a(\boldsymbol x)$ are positive definite with probability tending to one on the evaluation sample, so that the matrix square roots and corresponding Fr\'echet derivatives are well defined. In addition, $\widehat w$ is uniformly bounded.
\end{assumption}

Assumption~\ref{assump:nuisance} is the standard product-rate condition for orthogonal estimation and is satisfied, for example, when each nuisance estimator converges faster than $n^{-1/4}$. We state a common rate for the nuisance estimators for simplicity; the underlying requirement is that the relevant products of estimation errors are negligible at the root-$n$ scale. Thus, the theory does not require correctly specified parametric models or root-$n$ nuisance estimation. This allows flexible regression and density-ratio methods while retaining regular inference for the low-dimensional robust target.

\begin{equation}
\mathcal R_\lambda^*=q(\boldsymbol\Gamma_\lambda^*)+\lambda\|\boldsymbol\Gamma_\lambda^*\|_F^2,
\qquad
\widehat{\mathcal R}_\lambda=\widehat q(\widehat{\boldsymbol\Gamma}_\lambda)+\lambda\|\widehat{\boldsymbol\Gamma}_\lambda\|_F^2.
\label{eq:Rls}
\end{equation}

\begin{theorem}[Consistency]\label{prop:consistency}
Under Assumptions~\ref{assump:pd-var}--\ref{assump:nuisance}, for every fixed $\lambda\ge0$,
\[
\widehat{\boldsymbol\beta}_\lambda\xrightarrow{p}\boldsymbol\beta_\lambda^*,\qquad
\widehat{\mathcal R}_\lambda\xrightarrow{p}\mathcal R_\lambda^*.
\]
\end{theorem}

The consistency result does not eliminate the underlying lack of identification. Even if the nuisance functions were known, the cross-world coupling would remain unidentified and the target would still be the maximin value over $\mathcal G$. Thus, the theorem establishes consistency for the proposed robust target rather than for a point-identified cross-world model.

\begin{theorem}[Inference for the ridge-stabilized robust surrogate utility]\label{thm:Q-AN}
Under Assumptions~\ref{assump:pd-var}--\ref{assump:nuisance}, for fixed $\lambda>0$,
\begin{equation}
\sqrt n(\widehat{\mathcal R}_\lambda-\mathcal R_\lambda^*)\xrightarrow{d}\mathcal N(0,\sigma_{\lambda,\mathcal R}^2),
\qquad
\sigma_{\lambda,\mathcal R}^2=\operatorname{Var}\{\psi_{\lambda,\mathcal R}^{s}(O)\}
+\kappa^{-1}\operatorname{Var}\{\psi_{\lambda,\mathcal R}^{t}(\boldsymbol X^{\mathrm{tar}})\}.
\label{eq:ANRlambda}
\end{equation}
Let $\widehat\sigma_{\lambda,\mathcal R}$ be consistent for $\sigma_{\lambda,\mathcal R}$ and let $\widehat\zeta_p$ consistently estimate the $p$th quantile of the Gaussian-norm term arising from estimation of the ridge contribution. With
\begin{align}
U_{n,\lambda,\alpha}
&=\widehat q(\widehat{\boldsymbol\Gamma}_\lambda)
+\{z_{1-\alpha/2}\widehat\sigma_{\lambda,\mathcal R}+\widehat\zeta_{1-\alpha/2}\sqrt{1+d}\}/\sqrt n,\nonumber\\[-2pt]
L_{n,\lambda,\alpha}
&=\widehat q(\widehat{\boldsymbol\Gamma}_\lambda)-\lambda(1+d)
-z_{1-\alpha/2}\widehat\sigma_{\lambda,\mathcal R}/\sqrt n,
\label{eq:UandL}
\end{align}
we have
\begin{equation}
\mathbb P\{\mathcal R^*\le U_{n,\lambda,\alpha}\}\ge1-\alpha+o(1),\qquad
\mathbb P\{\mathcal R^*\ge L_{n,\lambda,\alpha}\}\ge1-\alpha+o(1).
\label{eq:R*CI}
\end{equation}
\end{theorem}

The Gaussian limit concerns the ridge-stabilized robust surrogate utility $\mathcal R_\lambda^*$; Proposition~\ref{prop:ridge control} translates it into valid bounds for the original maximin surrogate utility $\mathcal R^*$. The result does not require the penalized optimizer to be interior because it concerns the optimal value rather than the optimizer. In particular, the ridge-stabilized utility can admit a regular first-order expansion even when the least-favorable coupling lies on the boundary of the ambiguity set, because inference for the optimal value does not require differentiability of the optimizer itself. Thus, inference for $\mathcal R_\lambda^*$, and consequently valid bounds for $\mathcal R^*$, are available under weaker conditions than inference for the individual coefficients of the learned transformation.

Since $\tau_\lambda^*=\mathcal R_\lambda^*/\overline D$, joint asymptotic linearity of $\widehat{\mathcal R}_\lambda$ and $\widehat{\overline D}$ also yields inference for the ridge-stabilized robust surrogacy measure. In particular, by the delta method,
\[
\sqrt n(\widehat\tau_\lambda-\tau_\lambda^*)
\xrightarrow{d}\mathcal N(0,\sigma_{\lambda,\tau}^2).
\]
In practice, we estimate $\sigma_{\lambda,\mathcal R}$ and $\sigma_{\lambda,\tau}$ by bootstrap resampling and use the resulting standard errors for Wald inference.

\begin{assumption}[Interior penalized optimizer]\label{assum:interior}
For the fixed ridge level $\lambda>0$, $\boldsymbol\Gamma_\lambda^*$ lies in the interior of $\mathcal G$.
\end{assumption}

Assumption~\ref{assum:interior} is needed only for inference on the transformation, not for inference on the robust surrogate utility. A boundary solution means that an ambiguity constraint is active, in which case small perturbations of the data can change the optimizer nonsmoothly and ordinary Wald inference for $\boldsymbol\beta_\lambda^*$ may fail. The utility-based conclusions of Theorem~\ref{thm:Q-AN} remain available in that situation.

\begin{theorem}[Asymptotic normality of the optimal transformation]\label{thm:beta-AN}
Under Assumptions~\ref{assump:pd-var}--\ref{assum:interior},
\[
\sqrt n(\widehat{\boldsymbol\beta}_\lambda-\boldsymbol\beta_\lambda^*)\xrightarrow{d}\mathcal N(\boldsymbol0,\boldsymbol\Sigma_\beta),
\qquad
\boldsymbol\Sigma_\beta=\operatorname{Var}\{\boldsymbol\psi_\beta^s(O)\}
+\kappa^{-1}\operatorname{Var}\{\boldsymbol\psi_\beta^t(\boldsymbol X^{\mathrm{tar}})\}.
\]
\end{theorem}

Theorem~\ref{thm:beta-AN} supports Wald inference for fixed linear contrasts of $\boldsymbol\beta_\lambda^*$ and accounts for uncertainty in estimating the least-favorable coupling. Detailed influence functions and variance estimators are provided in Appendix~\ref{app:beta-AN}.

Together, the results suggest a natural reporting hierarchy. The robust surrogate utility and its normalized surrogacy measure provide the main summaries of surrogate performance under counterfactual ambiguity. The estimated transformation describes how the surrogate is combined and can support scientific interpretation, but coefficient-level Wald inference additionally requires the interiority condition.

\section{Simulation studies}

We use simulation studies to examine three complementary aspects of the proposed framework. First, we evaluate the finite-sample accuracy and uncertainty quantification of the robust surrogacy measure under nonlinear outcome and surrogate models with covariate shift, with corresponding results for the learned robust transformation reported in the Appendix. Second, we examine how the proposed maximin procedure behaves when the observed data are compatible with counterfactual misalignment, focusing on whether both the learned transformation and the robust surrogate utility shrink toward zero while a working-independence analysis can yield misleading conclusions. Third, we assess the practical use of the learned transformation for predicting the treatment effect on the primary outcome in a future trial from the observed treatment effect on the surrogate. Unless otherwise stated, treatment is randomized with probability $0.5$, each experiment is repeated over 500 Monte Carlo replicates, and the ridge level is fixed at $\lambda=0.1$. Additional details of the data-generating mechanisms, nuisance estimation, and implementation are provided in Appendix \ref{sec:app:sim}.

\subsection{Finite-sample estimation and inference}

We begin by evaluating the finite-sample performance of the proposed estimator for the penalized robust surrogate utility $R_\lambda^*$ and its normalized version, the surrogacy measure, $\tau_\lambda^*=R_\lambda^*/\bar{D}$. We consider two settings with nonlinear outcome and surrogate models and covariate shift between the source and target populations. In Setting~1, the surrogate vector is highly informative for the treatment effect on the primary outcome after conditioning on baseline covariates. In Setting~2, we introduce a latent component $U$ into the primary outcome model so that the surrogate remains informative but does not fully capture treatment effect on the primary outcome. The full data-generating mechanisms for both settings are given in Appendix~\ref{app:61}. Results for the learned robust transformation are also reported in Appendix~\ref{app:61}.

\begin{table}[ht]
\centering
\begin{tabular}{rrrrrr}
  \hline
$n\!=\!N\!=\!1000$ & TRUE & EST & SD & ESE & CP \\ 
  \hline
   $R_{\lambda}$ & 14.738 & 14.521 & 1.323 & 1.308 & 0.954 \\  \hline
  $\bar{D}$ & 17.607 & 17.630 & 0.849 & 0.878 & 0.960 \\  \hline
    $\tau_{\lambda}$ & 0.837 & 0.824 & 0.064 & 0.061 & 0.948 \\  \hline
   \hline
   \hline
   $n\!=\!N\!=\!2000$ & TRUE & EST & SD & ESE & CP \\ \hline
     $R_{\lambda}$ & 14.738 & 14.607 & 0.897 & 0.871 & 0.950 \\ \hline
  $\bar{D}$ & 17.607 & 17.632 & 0.578 & 0.611 & 0.960 \\ \hline
    $\tau_{\lambda}$ & 0.837 & 0.828 & 0.041 & 0.038 & 0.968 \\ \hline
\end{tabular}
\caption{Finite-sample performance of the proposed estimator in Setting~1. TRUE denotes the population target, EST the mean of the estimates, SD the empirical standard deviation, ESE the average estimated standard error, and CP the empirical coverage probability of the nominal 95\% confidence interval. The table reports estimates for the penalized robust surrogate utility $R_{\lambda}$, its normalizing quantity $\bar{D}$, and the normalized robust surrogacy measure $\tau_{\lambda}=R_{\lambda}/\bar{D}$.}\label{set1}
\end{table}

\begin{table}[ht]
\centering
\begin{tabular}{rrrrrr}
  \hline
$n\!=\!N\!=\!1000$ & TRUE & EST & SD & ESE & CP \\   
  \hline
      $R_{\lambda}$ & 3.727 & 3.820 & 0.953 & 0.991 & 0.955 \\  \hline
  $\bar{D}$ & 10.055 & 10.394 & 0.789 & 0.787 & 0.937 \\  \hline
    $\tau_{\lambda}$ & 0.371 & 0.366 & 0.081 & 0.088 & 0.965 \\ 
   \hline
     \hline
    $n\!=\!N\!=\!2000$ & TRUE & EST & SD & ESE & CP \\   \hline
    $R_{\lambda}$ & 3.727 & 3.711 & 0.628 & 0.627 & 0.954 \\  \hline
  $\bar{D}$& 10.055 & 10.278 & 0.551 & 0.519 & 0.938 \\  \hline
  $\tau_{\lambda}$ & 0.371 & 0.360 & 0.051 & 0.054 & 0.962 \\  \hline
\end{tabular}
\caption{Finite-sample performance of the proposed estimator in Setting~2. TRUE denotes the population target, EST the mean of the estimates, SD the empirical standard deviation, ESE the average estimated standard error, and CP the empirical coverage probability of the nominal 95\% confidence interval. The table reports the estimates for the penalized robust surrogate utility $R_{\lambda}$, its normalizing quantity $\bar{D}$, and the normalized robust surrogacy measure $\tau_{\lambda}=R_{\lambda}/\bar{D}$.}\label{set2}
\end{table}

Across both settings, Tables~\ref{set1} and~\ref{set2} show that the proposed estimator has good finite-sample performance, with estimation accuracy and precision generally improving as the sample size increases from $n=N=1000$ to $n=N=2000$. In Setting~1, the estimates of $R_{\lambda}$, $\bar{D}$, and $\tau_{\lambda}$ exhibit small bias, and the empirical and estimated standard errors are similar. The corresponding coverage probabilities are close to the nominal 95\% level. Setting~2 is more challenging because of the additional latent component in the outcome treatment effect. Nevertheless, estimation of $R_{\lambda}$, $\bar{D}$, and $\tau_{\lambda}$ remains accurate, with coverage probabilities close to 95\%. Overall, the results support accurate estimation and reliable uncertainty quantification for the proposed robust surrogacy measures, with performance improving at the larger sample size. Results for the learned transformation coefficients, reported in Appendix~\ref{app:61}, show a similar improvement with increasing sample size.  

\subsection{Behavior under counterfactual misalignment}

We next examine how the proposed maximin procedure behaves when the observed data are compatible with counterfactual misalignment. We consider a single-surrogate design indexed by three parameters: the covariate signal $\sigma_X$, the strength of treatment-effect reversal $\rho_s$, and the magnitude of unexplained outcome variation $\sigma_U$. Smaller values of $\sigma_X$ correspond to weaker covariate information about cross-arm dependence, whereas larger values of $\rho_s$ and $\sigma_U$ create more challenging regimes in which the treatment effect on the surrogate can be poorly aligned with the treatment effect on the primary outcome. The full data-generating mechanism is given in Appendix~\ref{app:62}.

We compare the proposed DRO estimator with a working-independence benchmark, corresponding to the PTE-type analysis of \citet{wang2020surrogate}, and a counterfactual oracle that uses the true cross-arm dependence and serves only as an infeasible benchmark. For each method, we report the learned transformation coefficient $\hat\beta$ and evaluate the resulting transformation under the true counterfactual coupling through surrogate utility $R(\hat\beta,\Gamma_{\mathrm{true}})$. This common evaluation isolates the consequences of learning the transformation under different assumptions about the unidentified cross-arm dependence.

\begin{figure}[!ht]
    \centering
    \includegraphics[width=\textwidth]{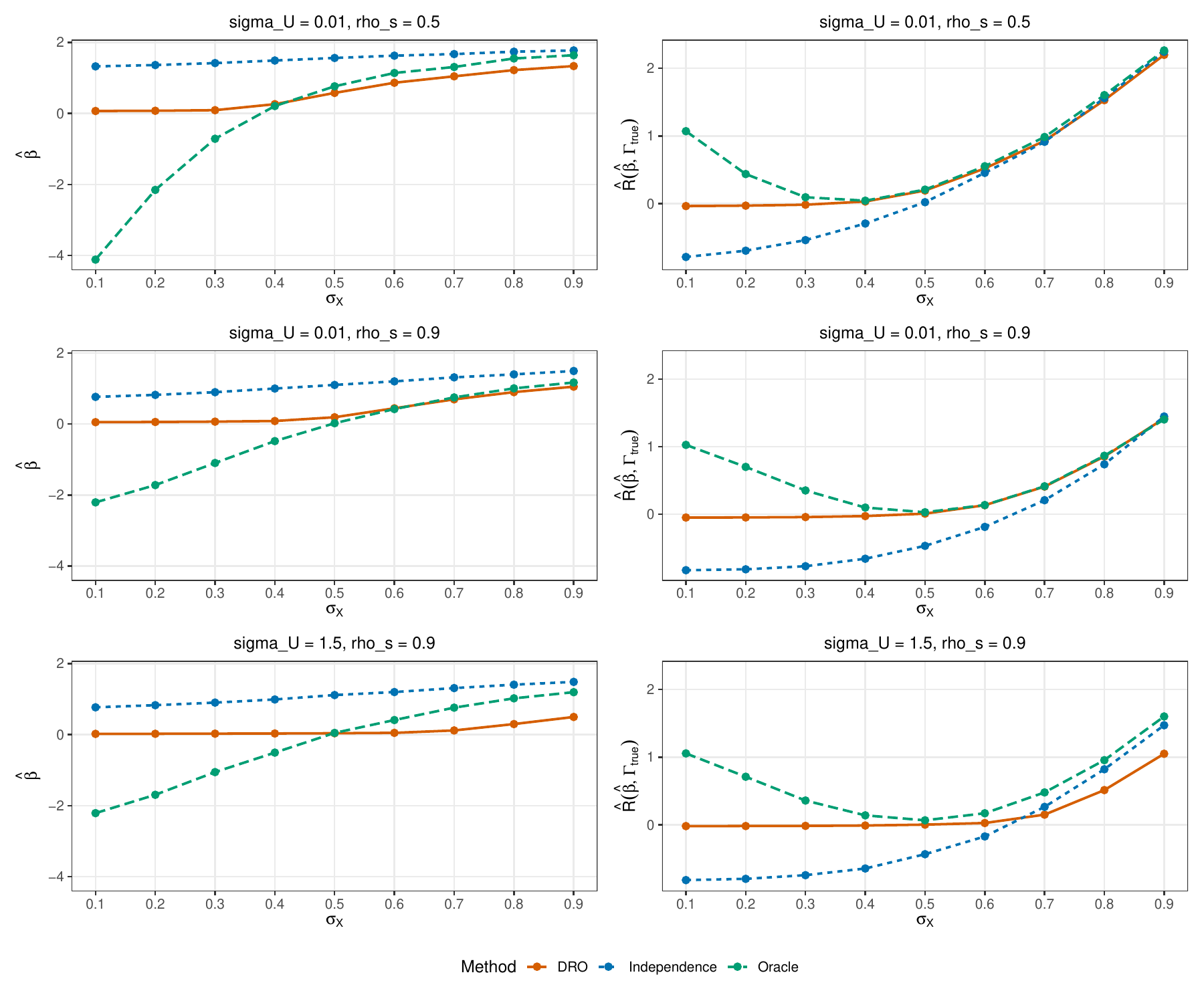}
    \caption{Behavior of the learned transformation and surrogate utility under counterfactual misalignment. 
    The left column shows the estimated transformation coefficient $\widehat{\beta}$, and the right column shows the corresponding explained-variance utility evaluated under the true counterfactual coupling, $R(\widehat{\beta},\Gamma_{\mathrm{true}})$, as functions of the covariate signal $\sigma_X$. 
    The three rows correspond to $(\sigma_U,\rho_s)=(0.01,0.5)$, $(0.01,0.9)$, and $(1.5,0.9)$, respectively. 
    We compare the proposed DRO estimator with the working-independence estimator and the infeasible counterfactual oracle.}
    \label{fig:paradox}
\end{figure}

Figure~\ref{fig:paradox} shows a clear contrast among the three procedures. When $\sigma_X$ is small, the working-independence transformation can have negative surrogate utility when evaluated under the true counterfactual coupling, indicating that it may perform worse than the null transformation. In contrast, the proposed DRO transformation remains close to zero, with its corresponding utility also near zero. This reflects the intended behavior of the maximin procedure: when the observed data remain compatible with counterfactual misalignment, the method avoids selecting a strongly nonzero transformation whose apparent usefulness depends on a favorable cross-world assumption. The oracle, which uses the true counterfactual dependence, can still achieve positive utility in these regimes and serves as a benchmark for the performance attainable if the cross-arm dependence were known.

As $\sigma_X$ increases, the covariate-driven component of cross-arm association becomes more informative, and the DRO transformation gradually moves away from the null. Its true-coupling utility correspondingly becomes positive and approaches the performance of the competing procedures. Stronger treatment-effect reversal, represented by larger $\rho_s$, delays this transition and extends the range of $\sigma_X$ over which the DRO transformation remains close to zero. Increasing $\sigma_U$ creates additional outcome variation not explained by the surrogate or observed covariates and makes the robust procedure more conservative in the corresponding regimes. Notably, the working-independence estimator can sometimes attain larger utility than DRO under the true coupling when the true coupling is favorable to the working-independence assumption; this does not contradict the maximin objective, which optimizes worst-case rather than true-coupling performance.

\subsection{Prediction of the treatment effect in a future trial}\label{prediction}

\begin{figure}[htb!]
    \centering
    \includegraphics[width=\textwidth]{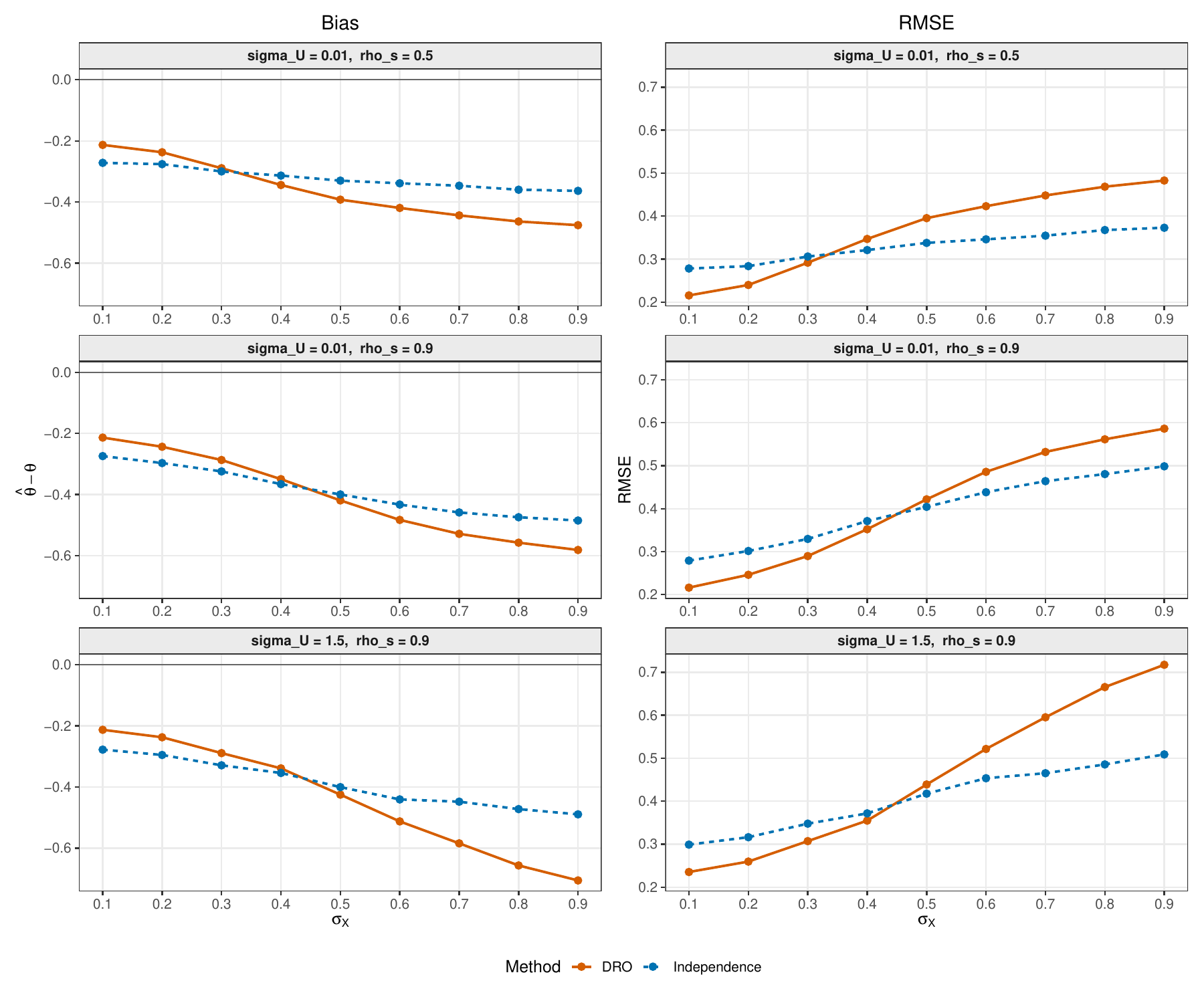}
    \caption{Prediction of the target-population treatment effect on the primary outcome using the learned surrogate transformation. The left column shows the Monte Carlo bias $\widehat\theta-\theta_Y$, and the right column shows the corresponding RMSE, as functions of the covariate signal $\sigma_X$. The three rows correspond to $(\sigma_U,\rho_s)=(0.01,0.5)$, $(0.01,0.9)$, and $(1.5,0.9)$, respectively. The proposed DRO estimator is compared with the working-independence estimator.}
    \label{fig:prediction}
\end{figure}

Our final experiment examines the downstream use of the learned surrogate transformation for predicting the treatment effect on the primary outcome in a future target trial. We use the same data-generating mechanism as in Section~6.2. The source trial is used to learn the surrogate transformation, while the target trial provides the observed treatment effect on the surrogate. Let $\widehat\Delta_{S,\mathrm{tar}}$ denote the estimated treatment effect on the surrogate in the target trial. For method $m\in\{\mathrm{DRO},\mathrm{WI}\}$, we predict the target treatment effect on the primary outcome by
\[
\widehat\theta_m=\widehat\beta_m\widehat\Delta_{S,\mathrm{tar}}.
\]
Because the primary potential outcomes are generated in the simulation, we can compare these predictions with the true target treatment effect
\[
\theta_{Y,\mathrm{tar}}
=
E\{Y^{(1)}-Y^{(0)}\mid L=1\}.
\]
The target primary outcome is used only to construct the simulation truth and is not used in forming the prediction. We summarize prediction performance using bias and root mean squared error (RMSE) over 500 Monte Carlo replicates. The same three representative regimes considered in Section~6.2 are shown in Figure~\ref{fig:prediction}.
 
The downstream prediction results complement the counterfactual-alignment experiment in Section~6.2. In regimes with weak covariate information or stronger misalignment, the DRO transformation is more conservative and can reduce prediction error relative to working independence by shrinking toward the null. As the covariate signal strengthens, the working-independence estimator can achieve smaller bias and RMSE because it is less conservative. Thus, the proposed procedure is not designed to uniformly dominate working independence in prediction under the realized coupling; rather, it trades some efficiency in favorable regimes for protection against adverse counterfactual dependence. Overall, the results illustrate how the robust transformation can be used for future-trial effect prediction while retaining the conservative behavior of the maximin construction under adverse counterfactual dependence.

\section{Real data analysis}
\label{sec:application}

In chronic kidney disease (CKD) trials, clinical endpoints such as kidney failure often occur late in the disease course, motivating the use of earlier surrogate outcomes to evaluate therapies intended to slow disease progression. Recent studies have identified decline in estimated glomerular filtration rate (GFR), commonly summarized by GFR slope, as a promising surrogate endpoint for CKD progression \citep{inker2019GFR,inker2023meta}. We apply the proposed distributionally robust framework to assess 18-month GFR change as a surrogate for two-year GFR change using randomized clinical trial data assembled by the Chronic Kidney Disease Epidemiology Collaboration (CKD-EPI\footnote{\href{https://www.tuftsmedicine.org/research-clinical-trials/research-institutes-research-specialty/chronic-kidney-disease-epidemiology-collaboration-ckd-epi}{https://www.tuftsmedicine.org/research-clinical-trials/research-institutes-research-specialty/chronic-kidney-disease-epidemiology-collaboration-ckd-epi}}), a collaborative research group focused on GFR measurement, estimation, and surrogate-endpoint evaluation in CKD.

We focus on four randomized clinical trials from CKD-EPI. The target trial is the MDRD-B study (sample size 255), a multicenter randomized trial among patients with more advanced CKD; in our analysis, the treatment contrast is very-low-protein diet supplemented with keto acids and amino acids versus low-protein diet \citep{beck1991design,klahr1994effects}. We consider three source trials. The first source trial is the MDRD-A study (sample size 584), a parallel randomized trial among patients with moderate CKD; here, the treatment contrast is low-protein diet versus usual-protein diet \citep{klahr1994effects}. The second is the DAPA-CKD trial (sample size 4,041), a randomized, double-blind, placebo-controlled trial of patients with CKD, with or without type 2 diabetes, evaluating the effects of dapagliflozin on kidney and cardiovascular outcomes \citep{heerspink2020dapagliflozin}. The third is the ALTITUDE trial (sample size 8,150), which was a randomized, double-blind, placebo-controlled trial evaluating the effects of aliskiren on cardiovascular and renal outcomes in patients with type 2 diabetes at high cardiorenal risk \citep{parving2012cardiorenal}.

The primary outcome $Y$ is the two-year change in GFR from baseline, and the surrogate $S$ is the 18-month change in GFR from baseline. The baseline covariates $\boldsymbol X$ include age, sex, baseline GFR, and baseline serum creatinine. We first assess surrogacy in the target trial using the proposed robust surrogacy measure learned from each source trial. We then examine the downstream use of the learned surrogate transformation for predicting the treatment effect on the primary outcome in the target trial when treatment assignment and the surrogate outcome, in addition to baseline covariates, are observed in the target trial. We compare predictions based on the proposed DRO transformation with those based on the working-independence (WI) transformation.

As in the simulation studies, we solve the DRO problem using ridge regularization with $\lambda=0.1$ and an operator-norm margin $\varepsilon_\Gamma=0.05$ for $\boldsymbol\Gamma$. Recall that $\widehat\Delta_{S,\mathrm{tar}}$ is the estimated treatment effect on the surrogate in the target trial. For $m\in\{\mathrm{DRO},\mathrm{WI}\}$, the surrogate-based prediction of the target treatment effect on the primary outcome is
\(
\widehat\theta_m=\widehat\beta_m\widehat\Delta_{S,\mathrm{tar}}.
\)
Because the two-year primary outcome is also observed in the target trial, we use the treatment effect $\widehat\theta_{Y,\mathrm{tar}}$ estimated directly from the randomized target-trial data as a benchmark for evaluating the surrogate-based predictions. Thus, the target primary outcome is used for evaluation rather than for constructing $\widehat\theta_m$.

\begin{table}[ht]
\centering
\begin{tabular}{llrrrrrr}
\hline
Source trial &  & $\tau_{\mathrm{DRO}}$ & $\beta_{\mathrm{DRO}}$ & $\beta_{\mathrm{WI}}$ & $\theta_{\mathrm{DRO}}$ & $\theta_{\mathrm{WI}}$ & $\theta_{Y,\mathrm{target}}$ \\
\hline
MDRD-A   & EST & 0.282 & 1.398 & 1.391  & 2.336  & 2.325   & 1.661 \\
         & ESE & 0.182 & 1.425 & 6.070  & 2.747  & 9.200   & 0.717 \\
\hline
DAPA-CKD & EST & 0.596 & 0.747 & 0.876  & 1.248  & 1.464   & 1.661 \\
         & ESE & 0.265 & 0.359 & 1.080  & 0.719  & 1.477   & 0.717\\
\hline
ALTITUDE&EST & 0.000 & -0.001 & 2.321 & -0.002 & 3.877 & 1.661 \\ 
&ESE & 0.253 & 1.154 & 4.875 & 2.173 & 8.419 & 0.717 \\ 
   \hline
\end{tabular}
\caption{Estimated robust surrogacy measure, surrogate transformations, and surrogate-based predictions of the treatment effect on the two-year GFR change in the MDRD-B target trial. Results are shown separately using MDRD-A, DAPA-CKD, and ALTITUDE as the source trial. $\tau_{\mathrm{DRO}}$ denotes the proposed robust surrogacy measure; $\beta_{\mathrm{DRO}}$ and $\beta_{\mathrm{WI}}$ denote the learned surrogate transformations under the distributionally robust and working-independence approaches, respectively; $\theta_{\mathrm{DRO}}$ and $\theta_{\mathrm{WI}}$ are the corresponding surrogate-based predictions of the target treatment effect; and $\theta_{Y,\mathrm{target}}$ is the treatment effect estimated directly from the MDRD-B target data.}\label{real}
\end{table}

Table~\ref{real} summarizes the estimated robust surrogacy measure, learned surrogate transformations, and corresponding surrogate-based predictions for the MDRD-B target trial. The estimated robust surrogacy measure is largest when DAPA-CKD is used as the source trial, with $\widehat\tau_{\lambda,\mathrm{DRO}}=0.596$, followed by MDRD-A with $\widehat\tau_{\lambda,\mathrm{DRO}}=0.282$, while the estimate based on ALTITUDE is essentially zero. The DRO and WI transformations are similar for MDRD-A and DAPA-CKD, whereas for ALTITUDE the DRO coefficient is shrunk nearly to zero while the WI coefficient remains positive. The downstream predictions reflect these differences. For MDRD-A, the DRO- and WI-based predictions are nearly identical, although both are estimated with substantial uncertainty. For DAPA-CKD, both predictions are relatively close to the treatment effect estimated directly in MDRD-B, with the WI prediction somewhat closer and the DRO prediction having a smaller estimated standard error. For ALTITUDE, the DRO prediction is close to zero, whereas the WI prediction is substantially larger and more variable. These results illustrate that the DRO procedure can substantially attenuate the learned surrogate transformation when its utility is sensitive to counterfactual dependence, while the WI approach may yield more accurate predictions in particular source--target pairs when its working assumption is favorable.

Overall, the CKD-EPI application illustrates how the proposed framework can be used to assess and transport surrogate information across clinical trials while accounting for uncertainty in cross-arm dependence. Compared to the working-independence approach, the DRO transformation is generally more attenuated when the surrogate--outcome relationship is sensitive to the unidentified counterfactual dependence structure. The procedure evaluates the learned transformation over the set of admissible cross-arm dependence structures while preserving the information identified from the observed treatment arms, thereby providing a more cautious basis for surrogate utilization across trials.
 
\section{Discussion}\label{sec:discussion}

We proposed a distributionally robust optimization framework for surrogate evaluation that explicitly addresses the fundamental non-identifiability of the joint counterfactual distribution. Rather than relying on unverifiable assumptions about the cross-arm dependence structure of the counterfactual outcome--surrogate pairs, our approach learns a surrogate transformation by maximizing its worst-case utility over an admissible class of counterfactual couplings and reports the corresponding robust surrogacy measure. This yields a conservative assessment of surrogate utility without committing to a single unverifiable cross-world dependence structure.

Our framework further uses baseline covariates to model treatment-specific conditional moments of the outcome--surrogate vector and to transport surrogate evaluation across populations. By integrating ideas from transfer learning and debiased machine learning, the learned surrogate evaluation can be transported to external target populations using only baseline clinical characteristics, without requiring observations of either the surrogate biomarker or the primary clinical outcome in the target cohort. This makes the framework applicable when only baseline covariates are available in the target population.
 
The conservativeness of our distributionally robust approach is determined by the choice of the ambiguity set for the unidentified cross-arm dependence. For simplicity and computational efficiency, we parameterize the residual cross-arm dependence by a constant matrix $\boldsymbol\Gamma$ and restrict it to a compact convex set $\mathcal G$ satisfying an operator-norm bound. The full operator-norm ball provides a broad default ambiguity set, while narrower choices can encode additional scientific knowledge about plausible counterfactual dependence. A broader ambiguity set provides protection against reliance on a particular unverifiable coupling, but may yield a more conservative surrogate assessment. When prior biological or clinical knowledge is available, $\mathcal G$ can be restricted accordingly, reducing conservativeness while retaining robustness over the remaining admissible dependence structures. 

The constant-$\boldsymbol\Gamma$ restriction can be further relaxed by allowing the residual cross-arm dependence to vary with baseline covariates through flexible nonparametric or neural network models. Such extensions would enlarge the class of admissible counterfactual dependence structures and allow greater flexibility in how residual dependence varies across covariate profiles. However, this increased flexibility also introduces significant computational and theoretical challenges. A nonparametric formulation would require optimization over an infinite-dimensional function space, while flexible finite-dimensional parameterizations such as neural networks could also lead to substantially more difficult optimization, requiring the admissible dependence constraint to hold uniformly over the support of the baseline covariates. Developing scalable algorithms and valid parameterizations for such functional uncertainty sets is an important direction for future research. More broadly, an important practical question is how to balance statistical robustness against conservativeness: while larger uncertainty sets offer stronger protection against specific unverifiable dependence assumptions, they may also produce overly conservative surrogate assessments. Designing uncertainty sets that are both scientifically informed and computationally tractable remains an important avenue for future investigation.

Future work will extend the proposed framework to accommodate time-to-event outcomes, longitudinal surrogate measurements, and dynamic treatment regimes. These extensions would broaden the scope of the framework to surrogate-evaluation settings with more complex outcomes, longitudinal data structures, and treatment strategies.

{\setstretch{1} 
\bibliographystyle{apalike}
\bibliography{ref}
}

\end{document}